\documentclass[superscriptaddress,twocolumn,showpacs,prl,floatfix]{revtex4}

\usepackage{color}
\usepackage{tabularx}
\usepackage{epsfig}
\usepackage{amsmath}
\usepackage{graphicx}

\begin{document}

\title{The Ising Spin Glass in dimension five : link overlaps}

\author{P. H.~Lundow}
\affiliation {Department of Theoretical Physics, Kungliga Tekniska h\"ogskolan, SE-106 91 Stockholm, Sweden}

\author{I. A.~Campbell}
\affiliation{Laboratoire Charles Coulomb,
  Universit\'e Montpellier II, 34095 Montpellier, France}

\begin{abstract}
  Extensive simulations are made of the link overlap in five
  dimensional Ising Spin Glasses (ISGs) through and below the ordering
  transition. Moments of the mean link overlap distributions (the
  kurtosis and the skewness) show clear critical maxima at the ISG
  ordering temperature. These criteria can be used as efficient tools
  to identify a freezing transition quite generally and in any
  dimension. In the ISG ordered phase the mean link overlap
  distribution develops a strong two peak structure, with the link
  overlap spectra of individual samples becoming very
  heterogeneous. There is no tendency towards a "trivial" universal
  single peak distribution in the range of size and temperature
  covered by the data.
\end{abstract}

\pacs{ 75.50.Lk, 05.50.+q, 64.60.Cn, 75.40.Cx}

\maketitle

We have studied the spin and link overlaps (defined below,
Eqs.~\ref{qtdef} and \ref{qltdef}) in some detail for cubic Ising Spin
Glasses (ISGs) with near neighbor interactions in dimension five. The
apparently perverse choice of dimension five, just below the ISG upper
critical dimension $d_{\mathrm{ucd}}=6$, has a number of
motivations. ISGs in this dimension have almost never been studied
through simulations, but there are precise and reliable estimates for
the inverse critical temperatures from High Temperature Series
Expansion (HTSE) calculations \cite{daboul:04}. In terms of numbers of
spins $N$ the sizes $L=4, 6$ and $8$ used here are equivalent to three
dimensional samples of size $L=10, 20$ and $32$, so from the point of
view of $N$ the present samples are "large". However, for the same $N$
equilibration to criticality and beyond is much faster than in lower
dimensions.

We first show that at the ordering temperature the moments of the mean
link overlap distributions show characteristic peaks, reflecting the
onset of correlated spin clusters.  This phenomenon can provide both
fundamental information on the ordering process and an efficient tool
for the accurate evaluation of critical temperatures, in any dimension
and for any interaction distribution.

Then in the ordered state, rather than measuring large numbers of
samples and averaging over the observables recorded, we take an
alternative approach. We study a limited number of samples ($32$ for
each size) and record explicitly the equilibrium spin overlap $P(q)$
and link overlap $Q(q_{\ell})$ spectra for each individual sample over
a succession of inverse temperatures. Examples of individual $P(q)$
distributions for ISGs have been exhibited in the literature, whereas
for the link overlaps only mean $\lbrack Q(q_{\ell})(L,\beta)\rbrack$
distributions
\cite{ciria:93,katzgraber:01,hed:07,contucci:07,leuzzi:08,janus:10}
but no individual sample $Q(q_{\ell})$ spectra have been published as
far as we are aware. It turns out that individual ISG samples in fact
show strongly idiosyncratic link overlap spectra which provide insight
into the controversial question of the structure of the SG ordered
phase.

The link overlap parameter \cite{caracciolo:90} in ISG numerical
simulations is the bond analogue of the intensively studied spin
overlap. In both cases two replicas (copies) $A$ and $B$ of the same
physical system are first generated and equilibrated; updating is then
continued and the "overlaps" between the two replicas are recorded
over long time intervals. The spin overlap at any instant $t$
corresponds to the fraction $q(t)$ of spins in $A$ and $B$ having the
same orientation (both up or both down), and the normalized overall
distribution over time is written $P(q)$. The link overlap corresponds
to the fraction $q_{\ell}(t)$ of links (or bonds or edges) between
spins which are either both satisfied or both dissatisfied in the two
replicas; the normalized overall distribution over time is written
$Q(q_{\ell})$.  The explicit definitions are
\begin{equation}
  q(t)=\frac{1}{N}\,\sum_{i=1}^{N} S_{i}^{A}(t)S_{i}^{B}(t)
  \label{qtdef}
\end{equation}
and
\begin{equation}
  q_{\ell}(t)=\frac{1}{N_{\ell}}\sum_{ij}S_{i}^{A}(t)S_{j}^{A}(t)S_{i}^{B}(t)S_{j}^{B}(t)
  \label{qltdef}
\end{equation}
where $N$ is the number of spins per sample and $N_{l}$ the number of
links; spins $i$ and $j$ are linked, denoted by $ij$.  We will
indicate means taken over time for a given sample by
$\langle\cdots\rangle$ and means over sets of samples by
$\lbrack\cdots\rbrack$.  The physical distinction between the
information obtained from $P(q)$ and $Q(q_{\ell})$ is frequently
illustrated in terms of a low temperature domain picture \cite{bokil:00}.

  The Hamiltonian is as usual
\begin{equation}
  \mathcal{H}= - \sum_{ij}J_{ij}S_{i}S_{j}
  \label{ham}
\end{equation}
with the interactions the symmetric bimodal ($\pm J$) or Gaussian
distributions normalized to $\langle J_{ij}^2\rangle=1$. We will quote
inverse temperatures $\beta = 1/T$. $\beta_c = 0.3925(35)$ for the
bimodal case and $\beta_c = 0.420(3)$ for the Gaussian from HTSE
calculations \cite{daboul:04}. Equilibration and measurement runs were
performed by standard heat bath updating (without parallel tempering)
on randomly selected sites. Each sample started at infinite
temperature and was gradually cooled until it reached its final
temperature. For temperatures near $T_c$ each sample saw at least
$10^7$ sweeps before reaching its final temperature. Then another
$10^7$ sweeps were run before any measurements took place. Normally
there were about 10 sweeps between measurements, though less for
higher temperatures and more for lower temperatures.  At each
temperature and sample we collected $4\cdot 10^5$, $6\cdot 10^5$ and
$5\cdot 10^6$ measurements for $L=8$, $6$ and $4$ respectively.

For the Gaussian ISG it has been shown \cite{katzgraber:01} that in
equilibrium
\begin{equation}
  \lbrack\langle q_{\ell}\rangle(L,\beta)\rbrack= 1-\lbrack\langle |U|\rangle(L,\beta)\rbrack/\beta
  \label{qlGauss}
\end{equation}
where $\lbrack\langle |U|\rangle(L,\beta)\rbrack$ is the mean energy
per bond. The Gaussian data presented here satisfy this equilibrium
condition over the full temperature range used; the bimodal samples
equilibrated faster than the Gaussian ones and were equilibrated for
as long times so we will consider that for present purposes effective
equilibration has been reached. However we will note elsewhere that
the condition Eq.~(\ref{qlGauss}) is necessary but not stringent
enough for true equilibration.

For the symmetric bimodal ISG there are simple rules on $\langle
q_{\ell}\rangle$.  If $p_{s}(\beta)$ is the probability that a bond is
satisfied,
\begin{equation}
  |U(\beta)| \equiv 1-2p_{s}(\beta)
\end{equation}
and a strict lower limit on $\langle q_{\ell}\rangle$ (uncorrelated
satisfied bond positions) is given by
\begin{equation}
  \lbrack\langle q_{\ell}\rangle(L,\beta)\rbrack  \geq p_{s}^2 + (1-p_{s})^2 - 2p_{s}(1-p_{s}) \equiv \lbrack U(L,\beta)^2\rbrack.
\end{equation}
In the high
temperature limit $|U|(\beta) \to \tanh(\beta)$ so
\begin{equation}
  \lbrack\langle q_{\ell}\rangle(L,\beta)\rbrack/\lbrack(1-|U(L,\beta)|/\beta)\rbrack \to  3.
\end{equation}
For a pure near neighbor ferromagnet (so with translational
invariance) $\lbrack\langle q_{\ell}\rangle(L,\beta)\rbrack \equiv
\lbrack U(L,\beta)^2\rbrack$ at all temperatures.

\begin{figure}
  \includegraphics[width=3.in]{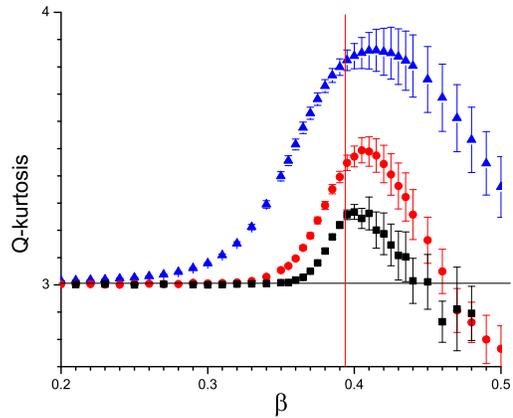}
  \caption{(Color online) The mean Q-kurtosis Eqn.~(\ref{Qkurt}) for
    the sets of bimodal samples with $L=4$ (blue triangles), $L=6$
    (red circles) and $L=8$ (black squares).}\protect\label{fig:1}
\end{figure}

In addition to the mean $\lbrack\langle q_{\ell}\rangle(\beta)\rbrack$,
important complementary information can be obtained from the moments of the
$Q(q_{\ell})$ distributions. The mean Q-kurtosis
can be defined by
\begin{equation}
  Q_{k} = \left\lbrack
  \left\langle \left(q_{\ell}-\langle q_{\ell}\rangle\right)^4\right\rangle \Big/
  \left\langle \left(q_{\ell}-\langle q_{\ell}\rangle\right)^2\right\rangle^2
  \right\rbrack
  \label{Qkurt}
\end{equation}
and the mean Q-skewness by
\begin{equation}
  Q_{s} = \left\lbrack
  \left\langle \left(q_{\ell}-\langle q_{\ell}\rangle\right)^3\right\rangle \Big/
  \left\langle \left(q_{\ell}-\langle q_{\ell}\rangle\right)^2\right\rangle^{3/2}
  \right\rbrack
  \label{Qskew}
\end{equation}
For the bimodal ISG the distributions are Gaussian at high
temperatures ($Q_{k}=3, Q_{s}=0$), with peaks in the mean
$Q_{k}(L,\beta)$ and $Q_{s}(L,\beta)$ at criticality, Figure 1.  The
amplitudes of the peaks decrease with increasing $L$; this could be
called an "evanescent" critical phenomenon as it will disappear in the
thermodynamic limit. Allowing for a weak finite size correction term,
the positions of the maxima $\beta_{max}$ for each set of peaks tend
accurately towards the HTSE $\beta_c$ with increasing $L$.  For the
Gaussian ISG an analogous peak behavior is observed.

Physically, the excess Q-kurtosis ("fat tailed" distributions) and
Q-skewness near $\beta_c$ in ISGs must be related to the build up of
inhomogeneous temporary correlated spin clusters around
criticality. The data show that they appear when the thermodynamic
limit correlation length ratio $\xi(\beta)/L$ is greater than some
value; then the Q-kurtosis and the Q-skewness each tend to a peak for
fixed $L$ as $\xi(\beta)$ diverges, so at $\beta_{c}$.  This simple
argument explains why in ISGs the peaks should be situated exactly at
$\beta_c$ in the large $L$ limit.  Indeed we will show elsewhere that
excess Q-kurtosis and Q-skewness peaks are not unique to ISGs;
analogous behavior with peaks tending to precisely $\beta_c$ with
increasing size is observed in a pure Ising ferromagnet, where the
data are not subject to sample averaging noise
\cite{lundow:12a}. Potentially Q-kurtosis and Q-skewness measurements
could be used as a standard procedure for estimating critical
temperatures. This method would not have the intrinsic disadvantages
of the usual crossing point techniques.

Turning to the ordered phase $\beta > \beta_c$, a longstanding
controversy concerns the correct physical description of ISGs for
finite dimensions \cite{parisi:83,fisher:86}. Generally the arguments
are focussed on the behavior to be expected at very large sizes and
very low temperatures. Unfortunately in practice the combination of
these limits renders direct observations inaccessible to numerical
simulations, so the conclusions must rely on extrapolations. Here our
aim is to reach a phenomenological description of an ensemble of
individual samples of limited size at temperatures somewhat below the
critical temperature, so in a region which can be studied directly by
simulations.
\begin{figure}
  \includegraphics[width=3.in]{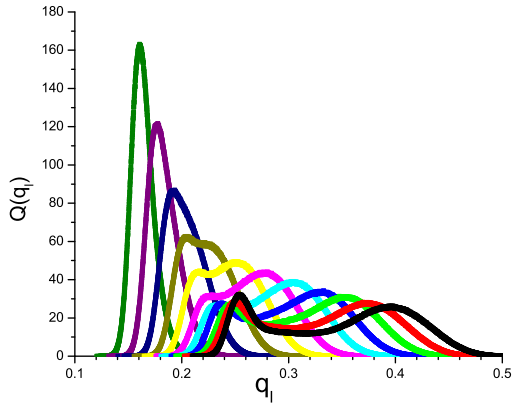}
  \caption{(Color online) The mean $L=6$ link overlap distributions
    $\lbrack Q(q_{\ell})(L,\beta)\rbrack$ for inverse temperatures
    $\beta=0.40, 0.42, 0.44, 0.46, 0.48, 0.50, 0.52, 0.54, 0.56, 0.58,
    0.60$ (from left to right; olive,purple,navy, brown, yellow, pink,
    cyan, blue, green, red, black).}\protect\label{fig:2}
\end{figure}

In Figure 2 we show mean bimodal link overlap spectra $\lbrack
Q(q_{\ell})\rbrack$ averaged over the $32$ sample set at $L=6$ and
$\beta$ up to $0.60$. Broadly similar double peak spectra have been
observed in other ISG measurements in dimensions $3$ and $4$
\cite{ciria:93,katzgraber:01,hed:07} but the present mean spectra
resemble even more closely mean field regime spectra in dimension $1$
with algebraically decaying interactions (Leuzzi {\it et al} Figure 2
\cite{leuzzi:08}). The presence of secondary peaks in mean $\lbrack
Q(q_{\ell})\rbrack$ spectra was explained by Bokil {\it et al}
\cite{bokil:00} as being due to the coexistence of a monodomain
configuration and a domain wall configuration, with the secondary peak
corresponding to one of the two replicas residing in one of these
configuration and the other replica in the other. By analogy with
results they had obtained on a Migdal-Kadanoff model, Bokil {\it et
  al} predicted that the secondary peak would melt rapidly into the
main peak with increasing $L$. This is not what we observe. As in the
mean field case the secondary peak becomes more and more cleanly
separated as the size is increased and as the temperature is
lowered. There appears to be no tendency towards "trivial" single peak
$\lbrack Q(q_{\ell})\rbrack$ structure in the range of $L$ and $\beta$
covered by the present data.

\begin{figure}
  \includegraphics[width=3.in]{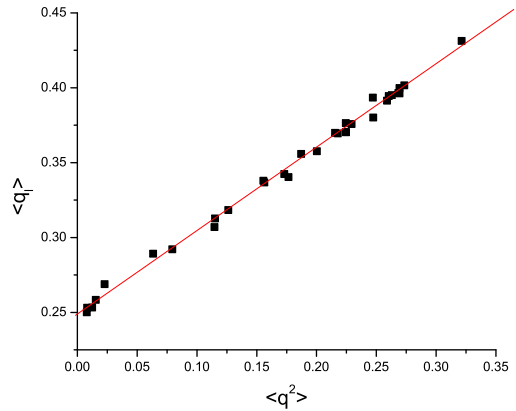}
  \caption{(Color online) A scatter plot of the link overlap $\langle
    q_{\ell}\rangle$ against the square of the spin overlap $\langle
    q^2\rangle$ for the set of $32$ bimodal samples with $L=6$ at
    $\beta = 0.60$.}\protect\label{fig:3}
\end{figure}
\begin{figure}
  \includegraphics[width=3.in]{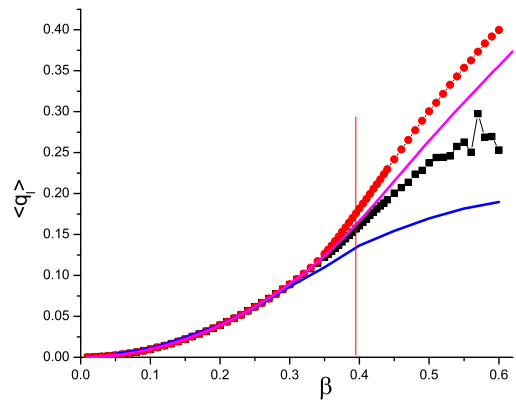}
  \caption{(Color online) The temperature dependence of the link
    overlap $\langle q_{\ell}\rangle(\beta)$ for two extreme $L=6$
    bimodal samples : sample $\# 22$ (red circles) and sample $\# 02$
    (black squares) together with the mean $\lbrack\langle
    q_{\ell}\rangle(\beta)\rbrack$ for the $L=6$ set (upper pink curve)
    and the square of the energy per link $\lbrack U(\beta)^2\rbrack$
    (lower blue curve). The vertical line indicates the HTSE critical
    $\beta{c}=0.3925$.}\protect\label{fig:4}
\end{figure}
\begin{figure}
  \includegraphics[width=3.in]{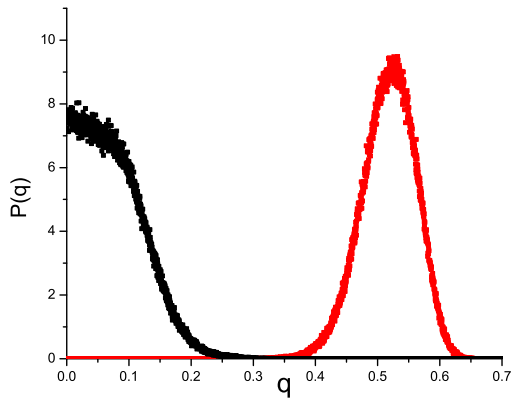}
  \caption{(Color online) The spin overlap distribution spectra
    $P(q)(\beta)$ for the $L=6$ samples $\# 22$ (red, right hand
    curve)and $\# 02$ (black, left hand curve) at $\beta=0.60$. (The
    spectra are symmetric about $q=0$).}\protect\label{fig:5}
\end{figure}
\begin{figure}
  \includegraphics[width=3.in]{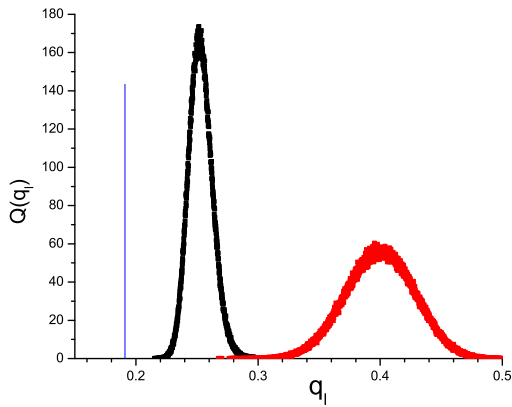}
  \caption{(Color online) The link overlap distribution spectra
    $Q(q_{\ell})(\beta)$ for the $L=6$ samples $\# 22$ (red, right hand
    curve) and $\# 02$ (black, left hand curve) at $\beta=0.60$. The
    vertical blue line is the strict random link limit $\langle
    q_{\ell}(\beta)\rangle = \lbrack U(\beta)^2\rbrack$.
  }\protect\label{fig:6}
\end{figure}

It is of interest to go into more detail and to examine the behavior
of the distributions $P(q)$ and $Q(q_{\ell})$ for individual samples
(see for instance \cite{yucesoy:12} for $P(q)$). Each of the $32$
realization sets can be taken as an unbiased sampling of the entire
population of possible realizations for a given size and temperature.
Only a couple of these spectra can be exhibited here.  In Figure 3 we
show a scatter pattern of $\langle q_{\ell}\rangle$ against $\langle
q^2\rangle$ for the $32$ sample set of $L=6$ individual bimodal
samples at the lowest temperature studied $\beta=0.60 \sim
1.5\beta_c$.  It can be seen that $\langle q^2\rangle$ and $\langle
q_{\ell}\rangle$ are correlated linearly as would be expected on the
grounds of overlap equivalence \cite{contucci:06}. We select two
extreme samples from this set, $\# 22$ with the highest and $\# 02$
with the lowest average $\langle q_{\ell}\rangle$. The temperature
dependence of $\langle q_{\ell}\rangle(\beta)$ for these samples
together with the mean $\lbrack\langle q_{\ell}\rangle(\beta)\rbrack$
and $[U(\beta)^2]$ are shown in Figure~\ref{fig:4}.  The degree of
localization of the satisfied links can be considered to be
represented by the difference between $\langle q_{\ell}\rangle(\beta)$
and the random link position limit $[U(\beta)^2]$.  It can be seen
that both samples begin to localize their links somewhat before
$\beta_{c}$; when the temperature is lowered further, the localization
in sample $\# 22$ is always about twice as strong as that of sample
$\# 02$. When the same procedure is followed for $L=4$ and $L=8$ or
for the Gaussian interaction samples, very similar results are
obtained.

Finally we show, Figures 5 and 6, individual $P(q)$ and $Q(q_{\ell})$
spectra for these extreme samples at $\beta = 0.60$. Both on the
Replica Symmetry Breaking (RSB) \cite{parisi:83} and the droplet
\cite{fisher:86} approaches one expects for individual spectra
principal Edwards-Anderson (EA) self-overlap peaks at $q = \pm q(EA)$)
and $q_{\ell} = q_{\ell}(EA)$, together with [RSB] or without
[droplet] secondary peaks. The high extreme sample $\# 22$ spectra
follow the simple droplet pattern and are "trivial", but remarkably
for the low extreme sample $\# 02$, where the $P(q)$ spectrum is
concentrated around $q=0$, the $Q(q_{\ell})$ spectrum shows
essentially no weight at an EA peak position anywhere near that of the
other spectrum. The entire $Q(q_{\ell})$ spectrum for this sample (and
others like it) is concentrated around the mean spectrum secondary
peak position in Figure 2, so in the sense of Bokil {\it et al} it is
"all wall". It is samples of this type which are contributing the most
strongly to the secondary peak in the mean $\lbrack
Q(q_{\ell})\rbrack$ spectra. The only plausible interpretation which
suggests itself for these spectra lacking EA self-overlap peaks would
appear to be in terms of a complex configuration space, with many
coexisting orthogonal configurations, in which $A$ and $B$ replicas
will almost never find themselves in the same configuration.

In conclusion, the characteristic critical link overlap moment
behavior shown here in the particular case of dimension 5 should be
observable, {\it mutatis mutandis}, at criticality in any dimension
and for any interaction distribution. Link overlaps can thus provide a
novel and powerful tool for studying the critical behavior of ISGs and
for making precise estimates of their critical temperatures.

Within the ISG ordered phase, the rich sample to sample heterogeneity
as seen through the individual link overlap spectra as well as through
spin overlap distributions can furnish a critical test of theoretical
models.  Mean distributions give only an incomplete view in the
ordered phase. The present data in dimension five, close to but below
the ISG ucd, show for some realizations of the interactions spin and
link overlap spectra which correspond to a simple mono-domain
description. For other realizations no self-overlap peaks are seen in
the spectra, indicating that the configuration space is much more
complex.

\end{document}